# COVID-19 Misinformation and Disinformation on Social Networks - The Limits of Veritistic Countermeasures

Andrew Buzzell
abuzzell@yorku.ca

**The COVID-19 pandemic has been the subject of a vast amount of misinformation, particularly in digital information environments, and major social media platforms recently publicized some of the countermeasures they are adopting. This presents an opportunity to examine the nature of the misinformation and disinformation being produced, and the theoretical and technological paradigm used to counter it. I argue that this approach is based on a conception of misinformation as epistemic pollution that can only justify a limited and potentially inadequate response , and that some of the measures undertaken in practice outrun this. In fact, social networks manage ecological and architectural conditions that influence discourse on their platforms in ways that should motivate reconsideration of the justifications that ground epistemic interventions to combat misinformation, and the types of intervention that they warrant. The editorial role of platforms should not be framed solely as the management of epistemic pollution, but instead as managing the epistemic environment in which narratives and social epistemic processes take place. There is an element of inevitable epistemic paternalism involved in this, and exploration of the independent constraints on its justifiability can help determine proper limits of its exercise in practice.**

The COVID-19 pandemic has been the subject of a vast amount of misinformation, particularly in digital information environments, and major social media platforms recently publicized some of the countermeasures they are adopting. This presents an opportunity to examine the nature of the misinformation and disinformation being produced, and the theoretical and technological paradigm used to counter it. I'll argue that the theoretical approach is based on a conception of misinformation as epistemic pollution that can only justify a limited and potentially inadequate response , and that some of the measures undertaken in practice outrun this. In fact, social networks manage ecological and architectural conditions that influence discourse on their platforms in ways that should motivate reconsideration of the justifications that ground epistemic interventions to combat misinformation, and the types of intervention that they warrant.

**Efforts by social networks to inhibit COVID-19 misinformation**

On March 16 Twitter announced that it is "broadening [its] definition of harm to address content that goes directly against guidance from authoritative sources of global and local public health information" (Gadde and Derella 2020), and will remove content it deems harmful. Facebook announced on March 25 that it would take steps to "...remove COVID-19 related misinformation that could contribute to imminent physical harm" (Clegg 2020), and for content that doesn't "...directly result in physical harm, like conspiracy theories about the origin of the virus" (ibid.), the platform will reduce distribution and add warnings. These anti-misinformation efforts focus on removing content that is deemed to be harmful and false, relying on internal moderation teams and algorithms for implementation, and external fact-checking organizations for guidance, such as those supported by the Google News Initiative (Mantzarlis April 2, 2020), or the partnership Facebook initiated with the Italian fact-checking organization Facta to monitor COVID-19 content on WhatsApp (Pollina April 2 2020). The sheer volume of misinformation has accelerated the reliance on algorithmic censorship tools, which are still brittle and prone to errors identifying pragmatic features of content, such as when for several days a substantial amount of legitimate content discussing COVID-19 was removed by Facebook's AI moderation system. (@alexstamos, March 17 2020).

**Fake News, Misinformation and Disinformation**

Efforts to combat misinformation typically focus on the extent to which targeted content is both false and harmful, and Levy (2018) argues such content can be described as epistemic pollution - which degrades the information environment and interferes with social epistemic practices that rely on trust. The pollution model encourages us to think about countermeasures in terms of cleanup that aims to restore the information environment to its natural state. This model nicely captures efforts by social networks to filter misinformation systematically, as well as Levy's suggested remedies, such as regulating and charging epistemic polluters.

There are important questions to be asked about the ethical and practical challenges this strategy faces, but here I want to focus on the problem of information that is deceptive, harmful, but not false, or at least not quite false enough that the fact-checking paradigm would capture it. This is a key feature of many forms of disinformation and propaganda (Derakhshan and Wardle 2017 call this "mal-information"), a category some COVID-19 misinformation belongs to, and one which can't be combatted directly with fact-checks and moderation tools operating within the pollution model. Some forms of disinformation work by polarizing the discourse on a topic, reframing it in ways that bring the conversation into what Nguyen (2018) calls epistemic echo chambers, where evidence that would correct misinformation is preemptively discredited. By exploiting features of individual cognition and the epistemic properties of echo chambers, this kind of disinformation is highly resilient to veritistic correction (Chan 2017).

The spread of misinformation on social networks has increasingly been the subject of public and scholarly attention, particularly where it has involved political processes such as elections. In late 2019 the Oxford Internet Institute found "... evidence of organized social media manipulation campaigns in 70 countries, up from 48 countries in 2018 and 28 countries in 2017" (Bradshaw and Howard 2019 p. 5), and found that these efforts have overwhelmingly been focussed on manipulating political attitudes. Analysis of social media manipulation often centres on so-called "fake news", as do discussions of appropriate policy responses, and the flood of misinformation surrounding COVID-19 is frequently conceptualized as a problem of fake news. Habgood-Coote (fortchcoming) argues that the conceptual imprecision of the term "fake news" functions to create an epistemic enclosure, limiting our theoretical resources in ways that re-enforce the authoritarian usage of the term to propagandize against legitimate sources of information, and that dull the critical tools required to combat it by obscuring distinctions we can make within the category. Fake news obscures an important distinction between misinformation and disinformation (Wardle 2019), and the diversity of functions it performs. Some fake news is propaganda generated by state and non-state actors for specific strategic ends, whereas others is produced for economic reasons, or even satire. Ordinary misinformation can be transformed into disinformation by hostile actors to further strategic aims. That the strategic aims frequently relate very indirectly to the content of the misinformation is a key feature of modern propaganda techniques - it might be of incidental interest to promote some particular false belief, where the goal is to undermine trust in some institution or hinder the effectiveness of some mode of action (Paul and Matthews 2016). Content-based blocking will miss a lot of effective disinformation unless it casts a net with my wider reach, beyond that of directly harmful and false material.

**Polarization, trust, and the social epistemic function of social networks**

One of the strategic goals of propaganda is to increase polarization, which creates epistemic collateral damage that is difficult to repair because it undermines networks of trust that connect experts and their research to broader audiences. The social turn in the study of epistemology emphasizes the extent to which epistemic labour is decentralized (Hardwig 1985, Goldberg 2011) and connected via chains of testimony maintained by processes which depend in part on trust. Misinformation can create primary and secondary damage to the epistemic environment. The primary damage is whatever specific harms result from the false claims, and the secondary damage is the disruption of social epistemic practices by overwhelming them, or undermining them, such by weakening trust in expertise that might be used to combat misinformation. As Milgram (2015) argues, our epistemic dependency is highly complex and inescapable - if the mechanisms that maintain the integrity of the transmission and maintenance of testimonial chains breaks down we don't have a backup.

It is this feature of social epistemic practices that Nguyen (2018) argues is responsible for the difficultly of escaping from echo chambers. An echo chamber occurs when the normal corrective testimonial processes are undermined, not because, as with an epistemic bubble, dissenting views are excluded, but because they are actively discredited. It's not that we don't encounter countervailing evidence, instead we are predisposed to dismiss it, because we don't trust the credentials of the source or we are unable to recognize legitimate disagreement because we have adopted what Nguyen describes as a disagreement-reinforcement mechanism - potential sources of evidence are preemptively discredited. In a process Begby (2020) calls "epistemic grooming", in this kind of unhealthy epistemic environment we are not only manipulated to believe some proposition, we also come to accept "... higher-order evidence about the credibility of a range of sources that would seek to contradict those propositions" (Begby 2020). In a bad epistemic environment, this evidence is misleading. Echo chambers are a useful theoretical perspective on the commonly observed paradox that in a world rich with information and connectivity, in some environments epistemic outcomes are increasingly poor.

In some echo chambers, disagreement-reinforcement mechanisms are supported by propaganda, especially polarizing propaganda (Sarts et al 2019) and undermining propaganda (Stanley 2015) which shift the evaluation context in which specific claims are received. When a politically neutral claim, such as that vaccines don't cause autism, or that COVID-19 is much more severe than the common flu, is reframed in partisan terms that connect the discourse to that in an echo chamber, the distorting features of the echo chamber inoculate us against contradictory evidence, invoking cognitive biases such as motivated reasoning that will alter extent to which audiences can effectively evaluate validity. In fact, exposure to factual information can have the opposite effect (Nyhan & Reifler, 2010, Chan 2017), making misperceptions worse, and where we might expect that countering misinformation with factual information would improve epistemic outcomes, we find the opposite. Even when fact-checking is made available online, it's circulation is highly constrained along partisan lines (Shin and Thorson 2017) As a result, propaganda that reframes a factual issue in partisan terms can alter the way we evaluate facts and claims of authority.

Fact checking is also difficult to perform when there is legitimate disagreement between experts. For example, as one epidemiologist observes on Twitter, experts working on COVID-19 are "... pilloried for updating their beliefs based on new information", and "... our models and predictions about the future trajectory of epidemic [are] deemed politically desirable or oppugnant" (@CT_Bergstrom 2020) . This is an example of disinformation that fits poorly in the pollution model, involving content that isn't false, and that doesn't mislead by covertly shifting or collapsing contexts, but instead explicitly promotes recontextualization that indirectly strengthens misinformation and frames the subject in ways that bring it into existing echo chambers. This undercuts the credibility of experts generally, and especially those experts who might be used to support anti-misinformation efforts. Twitter and Facebook seek to promote authoritative sources, but these sources can become grouped into "discourse coalitions" (Hajer 2002) and evaluated by their group membership instead of their legitimate authority. We do not usually interpret the advice of scientists in a partisan way, but in an increasing number of contested domains this is now the norm (Metze and Dodge 2016).

One of the themes of disinformation in the COVID-19 information environment has been to question scientific consensus on the origin of the virus, promulgating narratives that it originates from a bioweapon, disputing the country of origin, and spreading false information about the rate of infection (Institute for Strategic Dialogue 2020). In the US, organized disinformation campaigns have amplified content claiming that the virus is not as serious public health organizations claim, and that the high economic costs are a consequences welcomed by some political actors - messaging which has a dual effect. It directly contributes to misinformation by spreading false claims, but the higher level effect is to politicize the narrative, which in turn enlarges the disinformation attack surface by framing the topic as part of partisan discourse. Disinformation which amplifies disagreement about the seriousness of the virus, and the political valence and economic costs of the response to it, creates a meta-narrative that distorts the evaluation of factual claims about the pandemic even when it doesn't address them directly. As a result, countermeasures based on the pollution model are likely to be inadequate if they don't push back against this auxiliary disinformation, and moreover, by the time a specific false claim achieves enough circulation to receive censure, its tendency to distort can persist by way of mere exposure (Pennycook, Cannon, Rand 2017). This is particularly challenging for countermeasures that target a "tipping point" (Wardle 2019) for intervention, where the content is circulated widely enough that adding higher-order evidence for it's significance by challenging it isn't itself misleading, but early enough that intervention can successfully prevent harm. In summary, Facebook and Twitter aim to remove and label false claims, and to amplify authoritative sources, but our confidence that authoritative sources can correct misinformation might be misplaced in an unhealthy epistemic environment where there is disinformation that can make misinformation more resistant to correction.

**Beyond fact-checking - altering the conditions of narrative success**

Interventions in the way we interact with online information, such as "accuracy nudges" that require users to perform micro-tasks judging truthfulness before sharing content online have been shown to be effective in reducing the re-transmission of fake news, and also have demonstrated effectiveness at limiting the reproduction of false information about COVID-19 (Pennycook et al 2020). But this sort of intervention is difficult to implement, and might conflict with earlier findings by the same authors that the repetition of false claims on social media creates an "illusory truth effect" which increases our credences in them even when they are known to be false (Pennycook, Cannon, Rand 2017). It is not always practical to provide the information and in-frastructure to implement an intervention that shifts the mindset of users before they post content to social platforms, but this sort of alteration to the architectural features of digital information environments has the capacity to change kinds of discourse that gains influence.

Researchers have observed that the information ecology of social networks allows health misinformation to spread rapidly and to resist remediation, and one of the reasons for this is the powerful role that misleading narratives play in changing and reinforcing beliefs (Braddock and Dillard 2016). In a review of the problems facing science communicators attempting to improve the quality of online health information environments, researchers find that the narrative structure of misinformation makes it difficult to combat veritistically. What's distinctive about narrative is that it "...describes the cause-and-effect relationships between events...." (Dahlstrom 2014) and provides the audience with context that influences their evaluation of content. Social networks "... exhibit novel and powerful uses of narrative and pose numerous methodological challenges, requiring collaborative interdisciplinary approaches." (Caulfied at al 2019 p 53), and demand unique interventions that are real-time and which involve the development and distribution of compelling counter-narratives. One of the reasons narratives are so resistant to fact-checking is because they include signals about reliable domain authorities, connections to partisan discourse, and disagreement-reinforcement mechanisms such as evidential preemption that make them resistant to rebuttal by experts and fact-checks, and instead, health and science communication should ".. recognize the important research around virality in terms of both information dissemination and interaction with platform algorithms, and to design communication strategies that use these findings to their benefit." (Caulfield et al 2019 p 55). On March 12 the UK government announced an investment in public diplomacy to counter COVID-19 disinformation (Department for International Development 2020), with a mandate to develop and distribute just this sort of content to counter misleading narratives about COVID-19. Where the pollution model would motivate efforts to remove bad content from the information environment, an ecological model would recommend modifying the environment so that better content gains influence.

There is a tendency to worry that the support of particular narratives, and interference with the freedom of users to post and distribute content online amounts risks a form of epistemic paternalism that is objectionably illiberal when it goes beyond blocking illegal or directly harmful mater-ial. Moderation standards based on principles of harm and veritism have the attraction of capturing the sort of "shouting fire in a crowded theatre" cases where censure could be compatible with an otherwise anti-paternalist attitude, whereas the promotion of counter-narratives would not, and the middle ground, such as accuracy nudges have been the subject of significant controversy on precisely the grounds that they are epistemically paternalist - that they change our mind, without our consent. (Hurd 2016)

But worries about this kind of paternalism should be tempered by recognition that there is no particularly natural information ecology on social platforms. The circulation of information is highly mediated by human and algorithmic moderation, and the behavioural impact of interaction design and gamification. Faced with objections that nudging people to make behavioural changes is paternalist, Sunstein argues that the manipulation of choice architecture is ubiquitous (2015), and that there are no paternalism-free contexts, an argument which is particularly applicable to the digital information environment, where the user-facing architecture of the web and mobile applications is highly monitored and optimized to manipulate user behaviour, using methodologies such as split-testing. One way to combat the spread of misinformation and to reduce the effectiveness of disinformation is to explicitly experiment with architectural and environmental conditions that influence the kinds of narratives that gain traction, and attempt to establish and protect signals of domain expertise.

This requires more ambitious interventions by social networks than these fact-checking efforts, but in fact, some of the other actions undertaken to combat COVID-19 misinformation already outrun what the pollution model would warrant. User interface design, while highly contextual, can have significant influence on user behaviour and information consumption, and interventions that highlight authoritative information can have significant effects on user behaviour (Miklosik and Dano 2016). WhatsApp and Twitter have both made changes to the way the reproduction of content by users is throttled and labelled in the hope this will indirectly inhibit viral misinformation. Google, Facebook, and Twitter have adopted information design patterns that influence users to consume information form sources on COVID-19 which the platforms have made editorial decisions to promote. Both platform expertise (to determine which patterns generate influence and capture attention a given environment) and domain expertise (to determine which in-formation should be made more salient) is necessarily leveraged to implement these policies. The reintermediation of domain authorities in publication and distribution processes can support efforts to rebuild the social epistemic environment so that misleading narratives cannot flourish so easily.

Habgood-Coote questions the claim that "[t]raditional media is a 'bulwark' against the threat of misinformation..." (Habgood-Coote forthcoming p 16) on the grounds that this reproduces authoritarian narratives about mass media manipulation that have become part of the meaning of the term "fake news" - it invites us to distrust information on the basis of origin.. This might be a good reason to resist the adoption of the term "fake news", and the dialectic that it suggests between legitimate and illegitimate outlets, rather than focusing directly on content. However, it overlooks the social epistemic function of media institutions to maintain trust in the testimonial chains linking experts to the public, as what Goldberg calls remote monitors (Goldberg 2011). This function is now performed in new ways, with novel failure modes, by social platforms, and while for many years platforms rhetorically and legally promoted the view that they are mere conduits for the free flow of content, in fact, as the comprehensive and rapid response to COVID-19 misinformation demonstrates, they have the technical and institutional resources to perform the gatekeeping functions of traditional media. One of those functions is to prevent misinformation and disinformation from jumping from illegitimate sources into the mainstream discourse. In an information environment with rapid distribution and relatively little epistemic friction, disinformation can be "traded up the chain" (Marwick & Lewis, 2017) from niche social networks to mainstream outlets, acquiring apparent authority in the process.

Social network responses to COVID-19 misinformation have included efforts to remove epistemic pollution by way of algorithmic and human moderation, architectural modifications to alter the circulation and perception of some information, promotion of domain expertise, and active policing of influential accounts. These actions are part of program of epistemic environmental stewardship that social networks inevitably undertake, and which extends beyond the fact-checking paradigm. Platform features that modulate the gamification and monetization of content production and distribution are regularly altered with significant discourse effects. To the extent that they are understood at all, these effects are typically measured within the narrow parameters of the direct incentives of the platform operator, but they have significant social epistemic effects, as the COVID-19 case illustrates, that deserves greater theoretical and ethical attention. The editorial role of platforms should not be framed solely as the management of epistemic pollution, but instead as managing the epistemic environment in which narratives and social epistemic processes take place. There is an element of inevitable epistemic paternalism involved in this, and exploration of the independent constraints on its justifiability can help determine proper limits of its exercise in practice.

**References:**

@alexstamos (Alex Stamos) March 17 2020 https://twitter.com/alexstamos/status/1240060312520232961

Begby, E. (2020). Evidential preemption. Philosophy and Phenomenological Research.

@ CT_Bergstrom (Carl T. Bergstrom) March 26 2020 https://twitter.com/CT_Bergstrom/status/1243252341756669953

Bradshaw, S., & Howard, P. N. (2019). The global disinformation order: 2019 global inventory of organised social media manipulation. Project on Computational Propaganda. Chicago

Caulfield, T., Marcon, A. R., Murdoch, B., Brown, J. M., Perrault, S. T., Jarry, J& Rachul, C. (2019). Health misinformation and the power of narrative messaging in the public sphere. Canadian Journal of Bioethics 2(2), 52-60.

Chan, M. P. S., Jones, C. R., Hall Jamieson, K., & Albarracín, D. (2017). Debunking: A meta-analysis of the psychological efficacy of messages countering misinformation. Psychological science, 28(11), 1531-1546. Chicago

Clegg, Nick (March 25 2020) Combating COVID-19 Misinformation Across Our Apps Retrieved from https://about.fb.com/news/2020/03/combating-covid-19-misinformation/

Derakhshan, Hossein, and Claire Wardle. "Information Disorder: Definitions." AA. VV., Understanding and Addressing the Disinformation Ecosystem (2017): 5-12.

Dahlstrom, M. F. (2014). Using narratives and storytelling to communicate science with nonexpert audiences. Proceedings of the National Academy of Sciences, 111(Supplement 4), 13614-13620.

Department for International Development March 12 2020 "UK aid to tackle global spread of coronavirus 'fake news'" Retrieved from https://www.gov.uk/government/news/uk-aid-to-tackle-global-spread-of-coronavirus-fake-news

Gadde and Derella (March 16 2020) An update on our continuity strategy during COVID-19 Retrieved from https://blog.twitter.com/en_us/topics/company/2020/An-update-on-our-continuity-strategy-during-COVID-19.html

Goldberg, S. (2011). The division of epistemic labor. Episteme, 8(1), 112-125.

Habgood-Coote, Joshua (forthcoming). Fake News, Conceptual Engineering, and Linguistic Resistance: Reply to Pepp, Michaelson, and Sterken, and Brown. Inquiry: An Interdisciplinary Journal of Philosophy.

Hajer, M. A. (2002). Discourse coalitions and the institutionalization of practice: the case of acid rain in Great Britain. In Argument Turn Policy Anal Plan (pp. 51-84). Routledge.

Hardwig (1985). 'Epistemic dependence'. Journal of Philosophy.

Hurd, Heidi M. "Fudging Nudging: Why Libertarian Paternalism is the Contradiction It Claims It's Not." Geo. JL & Pub. Pol'y 14 (2016): 703.

Institute for Strategic Dialogue (2020) Covid-19 Disinformation Briefing No.1 Retrieves March 27 2020 from http://www.isdglobal.org/wp-content/uploads/2020/03/COVID-19-Briefing-Institute-for-Strategic-Dialogue-27th-March-2020.pdf

Levy, Neil. (2018)"Taking responsibility for health in an epistemically polluted environment." Theoretical medicine and bioethics 39.2 (2018): 123-141.

Mantzarlis, Alexios (April 2 2020) "COVID-19: $6.5 million to help fight coronavirus misinformation" Google Retrieved from https://www.blog.google/outreach-initiatives/google-news-initiative/covid-19-65-million-help-fight-coronavirus-misinformation/

Marwick, Alice, and Rebecca Lewis. (2017) "Media manipulation and disinformation online." New York: Data & Society Research Institute.

Metze, T., & Dodge, J. (2016). Dynamic discourse coalitions on hydro-fracking in Europe and the United States. Environmental Communication, 10(3), 365-379.
Chicago

Millgram, E. (2015). The great endarkenment: philosophy for an age of hyperspecialization. Oxford University Press.

Miklosik, A., & Dano, F. (2016). Search engine Optimisation and google answer box. Communication Today, 7(1), 82.Chicago

Nguyen, C. T. (2018). Echo chambers and epistemic bubbles. Episteme, 1-21.

Nyhan, B., & Reifler, J. (2010). When corrections fail: The persistence of political misperceptions. Political Behavior, 32(2), 303-330.

Paul, C., & Matthews, M. (2016). The Russian "firehose of falsehood" propaganda model. Rand Corporation, 2-7.

Pennycook, G., Cannon, T. D., & Rand, D. G. (2018). Prior exposure increases perceived accuracy of fake news. Journal of experimental psychology: general, 147(12), 1865.

Pennycook, G., McPhetres, J., Zhang, Y., Lu, J. G., & Rand, D. G. (2020). Fighting COVID-19 Misinformation on Social Media: Experimental Evidence for a Scalable Accuracy-Nudge Intervention. Psychological science, 0956797620939054.

Pollina, Elvira (April 2 2020) "Facebook launches fact-checking service on WhatsApp in Italy to fight coronavirus hoaxes" Reuters. Retrieved from Retrieved from http://www.reuters.com/article/2015/03/11/us-mideast-crisis-iraq-workshop-idUSKBN0M718T20150311.

Sarts, Janis, Bay, Sebastian, Šnore, Guna, Melnychuk, Jazlyn (2019), "Threats to Democracy: Technological Trends and Disinformation" in Disinformation and Digital Democracy in the 21st Century Retrieved from http://natoassociation.ca/wp-content/uploads/2019/10/NATO-publication-.pdf

Shin, J., & Thorson, K. (2017). Partisan selective sharing: The biased diffusion of fact-checking messages on social media. Journal of Communication, 67(2), 233-255.

Stanley, Jason (2015). How propaganda works. Princeton University Press

Sunstein, Cass R. (2019). Going to Extremes: How Like Minds Unite and Divide. Oxford University Press

Sunstein, Cass R. (2015) "The ethics of nudging." Yale J. on Reg. 32 (2015): 413.

Wardle, Claire, (2019) Understanding Information Disorder, First Draft, October 2019 https://firstdraftnews.org/wp-content/uploads/2019/10/Information_Disorder_Digital_AW.pdf?x76701